\documentstyle[aps,eqsecnum]{revtex}
\renewcommand{\mathbf}[1]{\mbox{\boldmath $#1$}}
\begin{document}
\preprint{}
\draft
\title{Hamiltonian formalism for the Oppenheimer-Snyder model}
\author{R. Casadio}
\address{Dipartimento di Fisica, Universit\`a di Bologna, and \\
I.N.F.N., Sezione di Bologna, \\
via Irnerio 46, 40126 Bologna, Italy}
\maketitle
\begin{abstract}
An effective action in Hamiltonian form is derived for 
a self-gravitating sphere of isotropic homogeneous dust.  
Starting from the Einstein-Hilbert action for barotropic perfect 
fluids and making use of the symmetry and equation of state of the 
matter distribution we obtain a family of reduced actions for two 
canonical variables, namely the radius of the sphere and its ADM 
energy, the latter being conserved along trajectories of the 
former.
These actions differ by the value of the (conserved) geodesic energy 
of the radius of the sphere which defines (disconnected) 
classes of solutions in correspondence to the inner geometry and
proper volume of the sphere.
By replacing the (fixed) geodesic energy with its expression in terms
of the Schwarzschild time at the surface of the sphere and treating
the latter as a further canonical variable we finally 
obtain an extended action which covers the full space of solutions.
Generalization to the (inhomogeneous) Tolman model is shown to be
straightforward.
Quantization is also discussed.
\end{abstract}
\section{Introduction}
In Ref.~\cite{hk} the Hamiltonian formalism was developed
for general classical barotropic fluids with step- and 
$\delta$-function discontinuities coupled to Einstein's gravity.
In particular, for the case with one step-function discontinuity
which separates a region filled with matter from an external
asymptotically flat vacuum, the canonical Hamiltonian was shown 
to consist of one volume contribution plus the Arnowitt-Deser-Misner
(ADM) energy which appears as a surface contribution
in the asymptotically flat region \cite{k}.
No particular symmetry was assumed for the matter distribution.
\par
Subsequently, in Ref.~\cite{h} this formalism was applied to the case
of a thin shell (with a $\delta$-function discontinuity) by assuming
the spherical symmetry from the beginning.
Then, adapting to this context a treatment invented for the spherically 
symmetric vacuum by Kuchar \cite{kuchar}, 
the action was further reduced to a form which contains only the physical 
degrees of freedom of the system, that is the shell radius and its 
ADM energy.
\par
A somewhat simpler example of discontinuous fluid is represented by
a sphere of homogeneous dust immersed in the vacuum.
This system is known as the Oppenheimer-Snyder (OS) model of
gravitational collapse and its Lagrangian equations of motion were 
solved long ago \cite{os}.
However, to our knowledge, a clean and complete derivation from first 
principles of an action in Hamiltonian form for the OS model which
covers the whole space of solutions is missing 
(see {\em e.g.} Refs.~\cite{lund,peleg,cv} and section~\ref{quantum}) 
and this is precisely the aim of the present notes.
The relevance of such issue is evident if one considers that the 
construction of a canonical form of the action is an essential
step towards the canonical quantization of a dynamical system
(this is actually the aim of the above mentioned papers
\cite{lund,peleg,cv}).
\par
The logical scheme we will pursue is essentially the same as in \cite{h}.
We start from the general formalism given in \cite{hk} and, after having 
identified all the properties of the matter source, we use such information 
to partially fix the form of the metric both inside and outside the sphere
of dust.
We do this in details for the case of homogeneous dust and show that
no new degree of freedom arises for the case of inhomogeneous dust with 
whatever fixed shape for the density profile (Tolman model \cite{tolman}).
The final reduction is gained by applying the transformations defined
in \cite{kuchar} to that part of the action corresponding to the outer
vacuum and by solving some of the constraints.
In so doing we will see the important role played by the known matching
conditions at the surface of the sphere from the point of view of 
Ref.~\cite{kuchar} (see Ref.~\cite{israel} for the standard treatment).
\par
The provisional outcome of our approach is an action in Hamiltonian form 
which describes the constrained dynamics of two canonical variables 
representing respectively the radius of the sphere in Schwarzschild 
coordinates and the ADM energy.
The latter is not assumed to be constant nor is dust proper energy.
However, the ADM energy is proved to be conserved along the trajectories 
of the radius determined by the super-Hamiltonian constraint. 
As a further consequence of the equations of motion, dust proper energy 
is conserved as well.
\par
So far the super-Hamiltonian constraint coincides with the classical 
counterpart of the Wheeler-DeWitt equation \cite{dw,w} recently used 
in Refs.~\cite{cv,cfv} (see also Ref.~\cite{lund,peleg} for a different
identification of variables).
One also notes that such constraint depends on the (discrete) parameter 
of the 3-curvature inside the sphere and the (continuous) comoving radial 
coordinate of the surface of the sphere.
These two parameters are invariants (or {\em perennials}, see {\em e.g.} 
\cite{per} and Refs.~therein) related to the conserved geodesic energy 
of the radius of the sphere. 
Therefore one concludes that the price to pay for the attained 
simplification is that the space of OS solutions to Einstein's equations
must be divided into classes of solutions with fixed geodesic energy.
Then each class can be treated as a separate dynamical system and is 
assigned an action in canonical form.
\par
Finally we introduce an extended action which describes all the 
OS solutions.
This is done by expressing the geodesic energy as a function of the
Schwarzschild time at the surface of the sphere and treating the latter
as a further canonical variable together with its conjugate momentum.
The final form of the action then contains six canonical variables
and one constraint, thus showing that the system possesses two physical
degrees of freedom.
Of course the transformation to Schwarzschild time is singular at 
the horizon, but we note that such singularity can be removed from 
the action by turning to Kruskal (or other regular) coordinates.
\par
In section~\ref{rev} we briefly review the relevant results obtained
in \cite{hk}.
In section~\ref{matter} we partially fix the metric and ADM foliation
of space-time by inserting all the information defining the matter
source.
In section~\ref{dyna} we derive and interpret the effective actions.
In section~\ref{quantum} we comment on the quantization of the model 
and compare to other approaches in the literature.
\par
We use units in which $8\,\pi\,G_N=c=1$, Greek indices run from $0$ to 
$3$ and Latin indices from $1$ to $3$.
\section{Hamiltonian for a step-discontinuous fluid}
\label{rev}
To start with, we assume that our space-time manifold $(\Omega,\mathbf{g})$ 
admits a smooth ($C^1$) foliation into space-like hypersurfaces $S_t$ where 
the parameter $t$ plays the role of time and ranges from $t_i$ to $t_f$.
Then, since we want to describe the dynamics of matter which fills only 
a compact region of space, it is natural to divide $\Omega$ into the 
space-time volume $\Omega_{(-)}$ swept by the matter trajectories and its
complementary $\Omega_{(+)}=\Omega\setminus \Omega_{(-)}$ which are 
separated by a time-like boundary $\Gamma$.
Further $\Omega_{(+)}$ has another time-like boundary $\Gamma_+$ which 
also bounds $\Omega$ as a whole.
Correspondingly $S_t$, for each value of $t$, is divided into an 
{\em internal\/} region $S_{(-)}=\Omega_{(-)}\cap S_t$ where matter 
has support and an {\em external\/} region $S_{(+)}=\Omega_{(+)}\cap S_t$ 
which is empty ($S_t=S_{(-)}\cup S_{(+)}$).
The interior $S_{(-)}$ has only one boundary $\partial S_{(-)}:=\Sigma
=\Gamma\cap S_t$, while the exterior is bounded by both $\Sigma$ and 
$\Sigma_+=\Gamma_+\cap S_t$, where the latter is a 2-surface which will 
eventually be pushed to spatial infinity.
\par
To be more specific, let us introduce a coordinate $r\ge 0$ on $S_t$ which 
is {\em adapted\/} to the boundaries \cite{hk} so that the origin $r=0$ is 
placed inside the matter distribution, $\Sigma_t$ is at fixed $r=r_s$ and 
$\Sigma_+$ at $r=r_+>r_s$.
\par
In Ref.~\cite{hk} matter is in the form of a {\em perfect fluid\/}, 
therefore its energy-momentum tensor can be written (in a set of 
coordinates $x^\mu=(t,r,x^2,x^3)$ which cover $\Omega$)
\begin{eqnarray}
T^{\mu\nu}=\sqrt{-g}\,\left[\left(e+p\right)\,u^\mu\,u^\nu
+p\,g^{\mu\nu}\right]
\ ,
\label{Tf}
\end{eqnarray}
where $e=e(x^\mu)$ is the density, $p=p(x^\mu)$ is the pressure
and
\begin{eqnarray}
u^\mu={dx^\mu\over d\tau}
\end{eqnarray}
is the 4-velocity of the fluid which satisfies the usual time-like 
condition $u_\mu\,u^\mu=-1$.
\par
For the case $e=p=0$ in $S_{(+)}$ for all values of $t$, 
it was shown in \cite{hk} that the action can be written in the
canonical form with Hamiltonian
\begin{eqnarray}
{\cal H}=\int_{S_t} dr\,dx^2\,dx^3\,\sqrt{q}\,\left(N\,H+N_i\,H^i\right)
-{1\over 2}\,\int_{\Sigma_+} dx^2\,dx^3\, Q^{tt}\,\gamma_{tt}
\ ,
\label{Ig}
\end{eqnarray}
where $\mathbf{q}$ is the 3-metric on $S_t$, $q$ its determinant,
$N$ the lapse function, $N_i$ the shift functions, $\mathbf{\gamma}$
the 3-metric on $\Gamma$ and $\mathbf{Q}$ the extrinsic curvature of 
$\Gamma$ \cite{adm}.
Further 
\begin{eqnarray}
H:= -{1\over\sqrt{-g}}\left(G^{\mu\nu}-T^{\mu\nu}\right)\,
u_\mu\,u_\nu
\ ,
\end{eqnarray}
is the {\em super-Hamiltonian},
\begin{eqnarray}
H^i:= -{1\over\sqrt{-g}}\left(G^i_\mu-T^i_\mu\right)\,
u^\mu
\ ,
\end{eqnarray}
are the {\em super-momenta\/} and $G_{\mu\nu}$ the Einstein tensor.
The surface term at $\Sigma_+$ in Eq.~(\ref{Ig}) is then proved to equal 
the ADM energy of the system \cite{k}.
\par
We are now ready to simplify the above expressions by making use
of the assumed symmetry and equation of state of the matter source.
\section{Symmetry and equation of state of matter}
\label{matter}
The particular matter configuration under study is isotropic,
therefore we can assume that $(S_t,\mathbf{q})$ is {\em spherically
symmetric\/}.
This allows us to give $r$ the meaning of a radial coordinate,
with $r=0$ as the centre of the sphere $S_{(-)}$ and take eventually
$r_+\to+\infty$.
To complete the set of spatial coordinates on $S_t$ in a way
which is adapted to the symmetry we simply supplement two angles 
$(\theta,\phi)$.
In these coordinates we can then write $e=e(r,t)$ and $p=p(r,t)$.
\par
The shift functions $N_\theta$ and $N_\phi$ can be set to zero and
the 3-metric $\mathbf{q}$ on $S_t$ can be written without
loss of generality
\begin{eqnarray}
d\sigma^2=\Lambda^2\,dr^2+R^2\,\left(d\theta^2+\sin^2\theta\,d\phi^2
\right)
\ ,
\label{sph}
\end{eqnarray}
where $\Lambda=\Lambda(r,t)$ and $R=R(r,t)$ are unspecified functions 
which do not depend on the angular coordinates and which we require 
to be smooth ($C^1$) across $\Sigma$ and to have the 
correct fall-off behaviours for $r=r_+\to+\infty$ \cite{kuchar}.
In particular we want $(S_t,\mathbf{q})$ to be asymptotically flat
(the cosmological constant is set to zero)
when its border $\Sigma_+$ is pushed to spatial infinity, thus
\begin{eqnarray}
\Lambda(r,t)=1+{m\over r}+{\cal O}(r^{-2})
\ ,
\end{eqnarray}
where $m$ is the ADM energy.
\par
So far only the spherical symmetry has been used,
however our matter source has more nice features.
The next step is to recall that the pressure vanishes for dust, 
$p(e)=0$, therefore Eq.~(\ref{Tf}) simplifies to
\begin{eqnarray}
T^{\mu\nu}=\sqrt{-g}\,e\,u^\mu\,u^\nu
\ .
\label{Td}
\end{eqnarray}
We observe in passing that, as a vector field, $\mathbf{u}(r,t)$ is 
singular at crossing points where the paths of two dust particles 
meet.
\par
We have still one last piece of information to put in, namely 
{\em homogeneity\/}.
This property can be formulated in the following way:
{\em i}) along the streamlines of the particles of dust (with tangent vector
$\mathbf{u}$) the density $e$ is a function of proper time only
\cite{only}, and
{\em ii})  the density does not change if we shift the point
of inspection perpendicularly to a streamline,
that is along any space-like vector $\mathbf{m}$ such that 
\begin{eqnarray}
u_\mu\,m^\mu=0
\ .
\label{orto}
\end{eqnarray}
More formally the above conditions are written as 
($e$ is a scalar density)
\begin{eqnarray}
{\cal L}_{\mathbf{u}}\,e= u^\mu\,\partial_\mu e=f(\tau)
\ \ \ {\rm and} \ \ \
{\cal L}_{\mathbf{m}}\,e= m^\mu\,\partial_\mu e=0
\ ,
\label{homo}
\end{eqnarray}
where ${\cal L}$ is the Lie derivative and $f$ a function of proper time
to be determined separately (see section~\ref{dyna}).
\par
Eq.~(\ref{homo}) could be considered as a set of two partial differential 
equations which specify $e$ as a functional of the metric $\mathbf{g}$ 
which enters through the condition (\ref{orto}) and normalizations.
It is instead more useful to assume the existence of a preferred reference 
frame in which
\begin{eqnarray}
{\cal L}_{\mathbf{u}}=\partial_t=N\,\partial_\tau
\ \ \ {\rm and}\ \ \  
{\cal L}_{\mathbf{m}}=\pm\partial_r
\ ,
\label{como}
\end{eqnarray}
where the lapse function $N(r,t)=N(t)$ for $0\le r<r_s$ and we have 
restricted $\mathbf{m}$ according to the spherical symmetry
(${\mathbf m}$ in general is a three dimensional unit vector).
Of course in these coordinates the density,
\begin{equation}
e=e(t)
\ ,
\label{ome}
\end{equation}
is a function of time only, as follows from the first equation in 
(\ref{homo}) and the definitions (\ref{como}).
\par
Since the coordinates $r$ and $t$ have to cover smoothly the whole 
space-time manifold $\Omega$, Eq.~(\ref{como}) implicitly requires 
that the vector fields $\mathbf{u}$ and $\mathbf{m}$ are regular 
everywhere in $\Omega$.
This rules out {\em a priori} all configurations of matter
trajectories with crossings between streamlines and
$(r,t,\theta,\phi)$ is a {\em comoving frame} in which $r$ labels 
the lines of flow of the particles in the fluid and $t$ 
parameterizes the position along each streamline.
Further, since $\partial_t$ is chosen perpendicular to $S_t$,
then $N_r(r,t)=0$ for $0\le r <r_s$, and the corresponding 
foliation of space-time inside the matter source is also called 
{\em time-orthogonal} \cite{co_orto}.
\par
The choice of Eq.~(\ref{como}) and homogeneity also determine the 
3-metric $\mathbf{q}_{(-)}$ in $S_{(-)}$ to be given by the unique 
maximally symmetric Robertson-Walker line elements \cite{weinberg}
\begin{eqnarray}
d\sigma^2_-=K^2\,\left[{dr^2\over 1-\epsilon\,r^2}+r^2\,\left(
d\theta^2+\sin^2\theta\,d\phi^2\right)\right]
\ ,
\label{rw}
\end{eqnarray}
where $\epsilon=0,\pm 1$ for respectively flat, spherical and hyperbolic
$S_{(-)}$, $K=K(t)$ is a function of time only and $0\le r<r_s$ with
$r_s<1$ for $\epsilon=1$ (we will see in section~\ref{dyna} the
dynamical meaning of $\epsilon$ and $r_s$).
\par
In $S_{(+)}$ the density $e=0$ and no comoving frame can be defined.
Therefore the 3-dimensional metric $\mathbf{q}_{(+)}$ cannot be 
{\em a priori} reduced to a form simpler than the one given in 
Eq.~(\ref{sph}).
However, at the surface $\Sigma$ one still has to enforce suitable
matching conditions such that $(S_t,{\mathbf q})$ is a $C^1$ metric 
manifold as prescribed.
\par
For the pull-back of the 4-dimensional metric ${\mathbf g}$ on $S_t$
to be $C^1$ one needs ${\mathbf q}_{(-)}$, Eq.~(\ref{rw}), and 
${\mathbf q}_{(+)}$, Eq.~(\ref{sph}), to satisfy
\begin{eqnarray}
&&\left.\Lambda\right|_+=
\left.{K\over\sqrt{1-\epsilon\,r^2}}\right|_- 
\nonumber \\
&&\left.\dot\Lambda\right|_+=
\left.{\dot K\over\sqrt{1-\epsilon\,r^2}}\right|_- 
\nonumber \\
&&\left.\Lambda'\right|_+=
\left.{K\,\epsilon\,r\over 1-\epsilon\,r^2}\right|_-
\label{mc1}
\end{eqnarray}
and
\begin{eqnarray}
&& \left.R\right|_+=\left.K\,r\right|_- 
\nonumber \\
&&\left.\dot R\right|_+=\left.\dot K\,r\right|_- 
\nonumber \\
&& \left.R'\right|_+=K 
\ ,
\label{mc2}
\end{eqnarray}
where we have introduced the notations
$\left.A\right|_\pm:=\lim\limits_{\delta\to 0}\,A(r_s\pm\delta,t)$,
$\dot A:=\partial_t A$ and
$A':=\partial_r A$ for any function $A$.
\par
In the above we have assumed ${\mathbf g}$ is written in the usual 
ADM form with lapse function $N$ and shift functions $N_i$ \cite{adm}.
Of the latter only the radial one, denoted by $N_r$, survives
after the imposition of the spherical symmetry.
Moreover, we have just seen that in $S_{(-)}$ one has $N=N(t)$ because 
of homogeneity and $N_r=0$ because $S_{(-)}$ is a time-orthogonal 
foliation.
Thus, on $\Sigma$
\begin{eqnarray}
&& \left.N\right|_+=N(t) 
\nonumber \\
&&\left.\dot N\right|_+=\left.\dot N\right|_- 
\nonumber \\
&& \left.N'\right|_+=0
\label{mc3} 
\end{eqnarray}
and
\begin{eqnarray}
&& \left.N_r\right|_+=\left.\dot N_r\right|_+=\left.N'\right|_+=0 
\ .
\label{mc4}
\end{eqnarray}
The matching conditions (\ref{mc1}) and (\ref{mc2}) will turn out to 
be extremely powerful in simplifying the final form of the action.
\par
Let us finish the section with a quick look at the slightly more 
general case in which
\begin{eqnarray}
e=f(t)\,h(r)
\ ,
\label{ine}
\end{eqnarray}
in the time-orthogonal comoving frame (which is again supposed to exist).
This represents inhomogeneous dust with a fixed density profile that can 
only rescale in time and is the matter content of the solutions found by 
Tolman \cite{tolman}.
Their metric can be written as in Eq.~(\ref{rw}), but with $K=J(r)\,L(t)$ 
now being also a function of the position inside $S_{(-)}$,
\begin{eqnarray}
h=2\,{(r^2\,J)'\over J'\,J^2}
\ .
\end{eqnarray}
It is apparent that Eq.~(\ref{ome}) is a special case of 
Eq.~(\ref{ine}) with $h$ constant ($J\propto r$). 
Since $h(r)$ is fixed once and for all, there is indeed no more freedom
in this generalization with respect to the homogeneous case, thus
we will not consider explicitly the density distribution (\ref{ine})
in the following sections.
\section{Hamiltonian dynamics}
\label{dyna}
In the previous section the 3-metric inside the sphere of dust, 
$S_{(-)}$, has been uniquely fixed to the form in Eq.~(\ref{rw}) 
for $0\le r<r_s$ by making use of all the properties assumed for 
the matter source.
Only isotropy survives in the outer empty region, $S_{(+)}$, and 
this has led to the more generic form in Eq.~(\ref{sph})
for $r>r_s$ and to the set of matching conditions in 
Eqs.~(\ref{mc1})--(\ref{mc4}).
\par
Plugging in all the above information into the Hamiltonian in 
Eq.~(\ref{Ig}) gives the action
\begin{eqnarray}
I=I_{(-)}+I_{(+)}+I_{\Sigma_+}
\ .
\end{eqnarray}
The first (volume) contribution comes from $S_{(-)}$ and is given by
\begin{eqnarray}
I_{(-)}=12\,\pi\,V_s\,\int_{t_i}^{t_f} dt\,
\left(P_K\,\dot K-N\,H\right)
\ .
\end{eqnarray}
The ``parametric volume'' of the sphere
\begin{eqnarray}
V_s=\int_0^{r_s} {r^2\,dr\over\sqrt{1-\epsilon\,r^2}}
\end{eqnarray}
has been integrated out (along with a factor of $4\,\pi$ which we shall
always factor out for convenience) and
\begin{eqnarray}
H:=-{1\over 2}\,
\left({P_K^2\over K}+\epsilon\,K-{2\over 3}\,e\,K^3\right)
\label{h-}
\end{eqnarray}
is the internal super-Hamiltonian which contains dust proper energy,
\begin{eqnarray}
H_M:=V_s\,e\,K^3
\ ,
\label{hm}
\end{eqnarray}
and the momentum conjugated to $K$ is defined by
\begin{eqnarray}
P_K:=-{K\,\dot K\over N}
\ .
\label{pk}
\end{eqnarray}
\par
The second (volume) contribution for $S_{(+)}$ can be read 
straightforwardly from Ref.~\cite{kuchar},
\begin{eqnarray}
I_{(+)}=4\,\pi\,\int_{t_i}^{t_f} dt\,\int_{r_s}^{+\infty} dr\,
\left(P_\Lambda\,\dot\Lambda+P_R\,\dot R-N\,H_{(+)}-N^r\,H_r\right)
\ ,
\end{eqnarray}
with
\begin{eqnarray}
&&
H_{(+)}:=-{P_R\,P_\Lambda\over R}+{\Lambda\,P_\Lambda^2\over 2\,R^2}
+{R\,R''\over\Lambda}-{R\,R'\,\Lambda'\over\Lambda^2}
+{{R'}^2\over 2\,\Lambda}-{\Lambda\over 2}
\nonumber \\
&&
H_r:=P_R\,R'-\Lambda\,P_\Lambda'
\ ,
\end{eqnarray}
and
\begin{eqnarray}
&&
P_R:=-{R\over N}\,\left(\dot R-N^r\,R'\right)
\nonumber \\
&&
P_\Lambda:=-{\Lambda\over N}\,\left(\dot R-N^r\,R'\right)
-{R\over N}\,\left(\dot\Lambda-(N^r\,R)'\right)
\ .
\end{eqnarray}
As a very simple check one can see that from Eqs.~(\ref{mc1}) and 
(\ref{mc2}) it follows that $\left.H_r\right|_+=0$ and
\begin{eqnarray}
\left.H_{(+)}\right|_+={1\over 3\,V_s}\,
{r_s^2\over\sqrt{1-\epsilon\,r_s^2}}\,
\left(\left.H_{(-)}\right|_--3\,e\,K^3\right)
\ ,
\end{eqnarray}
so that the super-Hamiltonian encodes the (step-function) discontinuity
at $r=r_s$.
\par
Finally, the (surface) contribution 
\begin{eqnarray}
I_{\Sigma_+}=-4\,\pi\,\int_{t_i}^{t_f} dt\,N_+\,m
\ ,
\end{eqnarray}
contains the ADM energy and the lapse function $N_+:= N(r_+,t)$.
\par
The internal action $I_{(-)}$ is already in its simplest possible
form.
The external action $I_{(+)}$ instead can be greatly simplified
by the following change of canonical variables \cite{kuchar}.
We define a new momentum conjugated to $R$:
\begin{eqnarray}
\bar P_R:=P_R-\left({1\over F_+\,F_-}\right)\,{\Lambda\,P_\Lambda\over 2\,R}
-{R\over 2}\,\left(\ln\left|{F_+\over F_-}\right|\right)'
\ ,
\end{eqnarray}
where $F_\pm:={R'/\Lambda}\pm{P_\Lambda/ R}$, and introduce two new 
canonically conjugated variables,
\begin{eqnarray}
&&
M:={P_\Lambda^2\over2\,R}-{R\,{R'}^2\over2\,\Lambda^2}+{R\over 2}
\nonumber \\
&&
P_M:={\Lambda\,P_\Lambda\over R\,F_+\,F_-}  
\ .
\end{eqnarray}
The difference between the corresponding Liouville forms \cite{kuchar}
\begin{eqnarray}
\Delta:=
\bar P_R\,\delta R+P_M\,\delta M-P_R\,\delta R-P_\Lambda\,\delta\Lambda
=\delta(\omega)+\left(\delta R^2\,\ln\left|{F_+\over F_-}\right|\right)'
\ ,
\end{eqnarray}
once integrated from $r=r_s$ to $r=r_+\to +\infty$ contributes
at most a surface term at $r=r_s$ ($\delta(\omega)$ can be discarded
since it corresponds to boundary terms at $t=t_i$ and $t=t_f$
which do not change the equations of motion).
By applying the matching conditions (\ref{mc1}), (\ref{mc2})
it is easy to see that
\begin{eqnarray}
\int_{r_s}^{+\infty} dr\,\Delta=
-\left.R^2\,\delta\ln\left|{F_+\over F_-}\right|\right|_+
=\Phi(K,\dot K)\,\delta\dot K
\ .
\end{eqnarray}
Therefore $\Delta$ does not contribute to $I_{(+)}$
and we can map $(R,P_R;\Lambda,P_\Lambda)$ to $(R,\bar P_R; M,P_M)$
without getting any new term in the action.
We point out that this is a major consequence of the smoothness of
$S_t$ across $\Sigma$ \cite{israel}.
\par
In $S_{(+)}$ the new momentum conjugated to $R$ and $M'$ are
linear combinations of the constraints $H_{(+)}$ and $H_r$
\cite{kuchar},
\begin{eqnarray}
&&
\bar P_R={1\over F_+\,F_-}\,\left({P_\Lambda\over R}\,H_{(+)}
+{R'\over\Lambda^2}\,H_r\right)
\nonumber \\
&&
M'=-{1\over\Lambda}\,\left(R'\,H_{(+)}+{P_\Lambda\over R}\,H_r\right)
\ ,
\end{eqnarray}
and vanish along classical solutions.
Then one can substitute $I_{(+)}$ with
\begin{eqnarray}
\bar I_{(+)}=4\,\pi\,\int dt\,\int_{r_s}^{+\infty} dr\,\left(
P_M\,\dot M+\bar P_R\,\dot R-N^R\,\bar P_R-N^M\,M'\right)
\ ,
\end{eqnarray}
where the new Lagrange multipliers $N^R$ and $N^M$ can be expressed
in terms of $N$ and $N^r$ \cite{kuchar}.
\par
At this point we are ready to reduce the action.
First we solve the constraints 
\begin{eqnarray}
\bar P_R=0\ ,
\ \ \ \ \ \ M'=0
\end{eqnarray}
and substitute the solutions back into $\bar I_{(+)}$.
Then we rename $N_+:=-\dot T$ and get the reduced action
\begin{eqnarray}
I&=&
12\,\pi\,V_s\,\int_{t_i}^{t_f} dt\left(P_K\,\dot K-N\,H\right)
+4\,\pi\,\int_{t_i}^{t_f} dt\,\left(\dot M\,\int_{r_s}^{+\infty} 
dr\,P_M +\dot T\,m\right)
\nonumber \\
&=&
12\,\pi\,V_s\,\int_{t_i}^{t_f} dt\left[P_K\,\dot K+p\,\dot m
+{N\over 2}\,\left({P_K^2\over K}+\epsilon\,K-{2\,m\over r_s^3}
\right)\right]
\ ,
\label{Ieff1}
\end{eqnarray}
where in the last line we have inferred from $M'=0$ that
\begin{eqnarray}
m:=\left.M\right|_+={r_s^3\over 2}\,\left({P_K^2\over K}
+\epsilon\,K\right)
\ ,
\end{eqnarray}
and this, together with $H=0$, relates the ADM energy to the matter
proper energy $H_M$, Eq.~(\ref{hm}), according to
\begin{eqnarray}
m={r_s^3\over 3}\,e\,K^3:={r_s^3\over3\,V_s}\,H_M
\ ,
\label{prop}
\end{eqnarray}
where $r_s^3/3\,V_s$ is the ratio between the proper volume of the
sphere and the volume of a flat sphere of equal radius \cite{stephani}.
We have also defined the momentum $p$ canonically conjugated to
$m$ as
\begin{eqnarray}
p:={1\over 3\,V_s}\,
\left(\int_{r_s}^{+\infty} dr\,P_M-T\right)
\ .
\end{eqnarray}
\par
The time has come to clarify the dynamical meaning of the two parameters 
$\epsilon$ and $r_s$ in the OS model. 
In order to do this, we define two new canonical variables 
$R_s:=r_s\,K$, $P_s:=P_K/r_s$ and rewrite the action (\ref{Ieff1}) 
as (we omit an irrelevant overall factor and make use of the 
arbitrariness of the lapse function)
\begin{eqnarray}
I_{r_s}^\epsilon(R_s,m;P_s,p)=
\int_{t_i}^{t_f} dt\left[
P_s\,\dot R_s+p\,\dot m
-N\,H_{r_s}^\epsilon\right]
\ ,
\label{Ieff}
\end{eqnarray}
with
\begin{equation}
H_{r_s}^\epsilon=-
{1\over 2}\,\left({P_s^2\over R_s}+\epsilon\,r_s^2\,R_s
-2\,m\right)
\label{Hred}
\ .
\end{equation}
The super-Hamiltonian constraint then becomes ($d\tau:=N\,dt$ is 
the proper time at the surface of the sphere)
\begin{eqnarray}
\left({d R_s\over d\tau}\right)^2=-\epsilon\,r_s^2+{2\,m\over R_s}
\ ,
\end{eqnarray}
which is the expected equation of radial geodesic motion of the 
radius of the sphere $R_s$ in the external Schwarzschild metric 
with mass parameter \cite{stephani} 
\begin{eqnarray}
m=\int_0^{R_s} e\,R^2\,dR
\ ,
\label{mint}
\end{eqnarray}
and geodesic energy ($T_s$ is the Schwarzschild time at the surface 
of the sphere),
\begin{equation}
E:=\left(1-{2\,m\over R_s}\right)\,{dT_s\over d\tau}
\ ,
\label{E}
\end{equation}
given by
\begin{eqnarray}
E^2=1-\epsilon\,r_s^2
\ .
\label{Egeo}
\end{eqnarray}
One observes that for $\epsilon=\pm 1$ the function 
$E(\epsilon,r_s)=E(\pm r_s^2)$ is continuous in $r_s$, but
for $\epsilon=0$ the value of $r_s$ is totally arbitrary, since 
$E(0,r_s)=1$, $\forall\,r_s>0$.
\par
In the expressions above the mass parameter is a positive quantity 
and the geodesic energy ranges from $-1$ to $+\infty$ (
$-1\le E<1\Leftrightarrow\epsilon=+1\Leftrightarrow H_M/m>1$ for 
bound orbits and 
$E\ge 1\Leftrightarrow\epsilon=0,-1\Leftrightarrow H_M/m\le 1$ 
for unbound orbits \cite{stephani}).
Thus the quantities $E$ and $m$ define separate classes 
$[E,m]$ of OS space-time solutions to the Einstein equations, 
each of which (provided is not empty) can be treated as one constrained 
dynamical system with canonical variables $R_s$ and $P_s$.
This implies that each trajectory belonging to a given class $[E,m]$  
is specified by the initial conditions (at $\tau=\tau_i$)
\begin{eqnarray}
\left\{\begin{array}{l}
R_s(\tau_i):=R_0=\left({3\,m\over e(\tau_i)}\right)^{1/3} 
\\
\left({d R_s\over d\tau}\right)_{\tau=\tau_i}:=V_0=
\pm\sqrt{{2\,m\over R_0}-1+E^2}
\ .
\end{array}\right.
\label{init}
\end{eqnarray}
From the second equation in (\ref{init}) it follows that, in each 
$[E,m]$, $V_0$ admits just two opposite values (if any) once $R_0$ 
is given.
Further, from the first equation in (\ref{init}) one sees that the 
choice of $R_0$ (or $V_0$) is equivalent to fixing the dust proper 
energy density $e(\tau_i)$ which has totally disappeared from the
action.
\par
In these notes, following the basic idea of fully employing the symmetry 
and equation of state of the matter source but no further information, 
we have ended up in considering the larger disconnected classes 
of space-time solutions defined solely by the geodesic energy $E$,
that is $[E]=\cup_{m>0}\,[E,m]$. 
Therefore in our approach the ADM mass and its conjugate momentum have
become canonical variables along with $R_s$ and $P_s$ and each trajectory
in a given class $[E]$ is specified by the initial conditions $R_0$, 
$V_0$, $m_0:=m(\tau_i)$ and $v_0:=\dot m(\tau_i)$.
\par
In each $[E]$ the allowed values of $R_0$ and $V_0$ are not uniquely 
related. 
However, by varying the action in Eq.~(\ref{Ieff1}) with respect to $p$
one obtains $\dot m=0$, thus $v_0=0$ and the ADM energy is conserved 
($m=m_0$) as a consequence of the equations of motion.
Then, from Eq.~(\ref{prop}), one deduces the conservation of matter 
proper energy as a consequence of the conservation of $m$.
At this point Eqs.~(\ref{init}) relate $R_0$ and $V_0$ according to the
subclass $[E,m_0]$.
\par
We also observe that Eq.~(\ref{prop}) does not depend on the internal 
geometry.
In fact, the independence of the ADM energy from $\epsilon$
(internal 3-curvature) is another well known property of the
OS model \cite{stephani} related to Birkhoff's theorem 
\cite{weinberg}.
\par
From the definition of $E$ in Eq.~(\ref{E}) and Eq.~(\ref{Egeo}) one
can rewrite the term $\epsilon\,r_s^2$ in the super-Hamiltonian (\ref{Hred}) 
as a function of the Schwarzschild time $T_s$ which becomes a canonical 
variable with conjugate momentum
\begin{equation}
P_T=-\left(1-{2\,m\over R_s}\right)^2\,{R_s\,\dot T_s\over N}
\ .
\end{equation}
The extended action thus obtained finally reads 
\begin{eqnarray}
I^e(T_s,R_s,m;P_T,P_s,p)=\int_{t_i}^{t_f} dt\left[
P_T\,\dot T_s+P_s\,\dot R_s+p\,\dot m
-N\,H^e\right]
\ ,
\label{Iext}
\end{eqnarray}
where the extended super-Hamiltonian is
\begin{equation}
H^e=-{1\over 2}\,\left(
{P_s^2\over R_s}-{R_s\,P_T^2\over(R_s-2\,m)^2}
+R_s-2\,m\right)
\ .
\end{equation}
Since no parameters but only six canonical variables appear in the 
expression of $H^e$, one concludes that $I^e$ describes all OS 
solutions in the phase space ${\cal F}=\cup_{E\ge-1}\,[E]$.
There is still one constraint, thus the system admits two physical 
degrees of freedom.
In this respect, the extended action $I^e$ and the super-Hamiltonian 
$H^e$ represent the main results of these notes.
\par
Let us finish the section with two remarks.
The reduced action $I_{r_s}^\epsilon$ is regular for all values of 
$R_s$.
On the contrary, since $T_s$ diverges at the horizon, the extended 
action is singular for $R_s=2\,m$. 
However one can eliminate such singularity by changing from 
the Schwarzschild coordinates $R_s$ and $T_s$ to the Kruskal 
(or other regular and possibly more convenient) coordinates 
of the surface of the sphere (we leave this analysis to future 
developments).
Finally, the generalization to the case of inhomogeneous dust defined 
by Eq.~(\ref{ine}) is straightforward and essentially amounts to 
changes in numerical coefficients appearing in the action.
\section{Remarks on quantization}
\label{quantum}
The main reason of interest in the canonical formalism of the OS model
relies in the hope that its quantization might supply a better 
understanding of the physics of gravitational collapse.
\par
Although very simple in its final form, our extended super-Hamiltonian 
constraint $H^e=0$ contains several difficulties along the path of 
quantization.
Not only the kinetic term leads to the problems of {\em operator ordering} 
and {\em Hermiticity} of $\hat H^e$ but, 
since $H^e$ has non-vanishing Poisson brackets with $T_s$, $R_s$
and their momenta,
the latter canonical variables cannot even be mapped into quantum 
observables straightforwardly.
Let us however assume that this can be solved by eliminating them 
in favour of related perennials (see \cite{per} and Refs.~therein),
as is already done for $T_s$ and $P_T$ when considering 
$H_{r_s}^\epsilon$, and focus on some aspects of the OS 
model treated in the literature.
\par
One of the first attempts to quantization is found in Lund \cite{lund}, 
where dust is represented by a homogeneous scalar field $\varphi$, 
which serves as a potential for the dust 4-velocity 
($u^\mu:=\partial^\mu\varphi$), and its conjugate momentum $P_\varphi$.
The scalar field, because of the normalization $u_\mu\,u^\mu=-1$,
has a matter super-Hamiltonian $H_M=P_\varphi$ and can be used to
define a time variable ($t:=-\varphi$) in order to turn the 
super-Hamiltonian constraint $H_G+H_M=0$ into a Schr\"odinger-like 
equation $[\hat P_\varphi+\hat H_G]\,\Psi=0$.
The latter does not suffer of the operator ordering ambiguity
but cannot be solved exactly.
The wavefunction $\Psi=\Psi(t,x)$ depends on the quantized canonical 
variable $x$ related to $R_s$ by a canonical transformation 
which also maps $H_G$ into the gravitational part ($H^\epsilon_{r_s}-H_M$) 
of the super-Hamiltonian in our reduced action (provided $m$ is fixed).
\par
The outer metric as well as the matching conditions with the 
inner metric at the surface of the sphere are never considered 
explicitly in \cite{lund} and the relation between the ratio  
$3\,V_s/r_s^3$ and the geodesic energy $E$ is thus missing.
Further, neither the ADM energy is discussed nor the ADM surface 
term is included so that the interpretation of the reduced action 
as describing (classes $[E,m]$ of) the OS model, rather 
than a cosmological model, is incomplete.
\par
On a different ground, one observes that the identification 
$t:=-\varphi$ prevents the second quantization of the matter content 
of the model and one cannot even count the number of dust particles.
This seems unsuitable for an astrophysical model in which one might 
rather prefer to study quantized matter on a (semi-)classical
background prior to quantization of gravity.
\par
More recently in Refs.~\cite{peleg,cv} the super-Hamiltonian in 
Eq.~(\ref{h-}) is used and the matching conditions at $r=r_s$ are 
correctly implemented, although somewhat {\em ad hoc}, with $m$
constant.
However the ADM surface term is again neglected and the partition
of the phase space into classes $[E,m]$ is not acknowledged.
Moreover, Peleg \cite{peleg} then goes back to the same action of
Ref.~\cite{lund}, in order to avoid the operator ordering ambiguity,
and solves the Schr\"odinger equation with dust as time in the
semi-classical (WKB) approximation.
\par
In Ref.~\cite{cv} the scalar field is not normalized and its
(second) quantization gives a proper energy $\langle\hat H_M\rangle$ 
proportional to the number of dust quanta in such a way that the
latter is not {\em a priori} constant. 
In this approach $K$ is treated semi-classically and plays the role 
of time (see Ref.~\cite{bfv} and Refs.~therein for the general 
formalism). 
\par
When the scalar field is not normalized one must be careful in 
asserting that it models dust since it is indeed equivalent to 
a perfect fluid with pressure equal to its Lagrangian density 
\cite{madsen}, $p\sim\frac{1}{2}\,(\dot\phi^2-\mu^2\,\phi^2)$.
However we observe that, if the scalar field has mass 
$m_\phi=\hbar\,\mu$, then the pressure oscillates with frequency
$\sim 2\,\mu$ and for $m_\phi\sim m_p\sim 10^{-27}$~kg (the proton
mass) this means a period $T\sim 10^{-23}$~s
(in the classical limit $\hbar\to 0$ with $m_\phi$ constant $T\to 0$).
It is thus reasonable to approximate the actual value of $p$ with
its average over one period (that is zero) provided $K$ does not 
change appreciably on the time scale $T$ 
({\em adiabatic approximation}).
In Ref.~\cite{cv} it was verified that in this approximation
($\dot R_s/R_s\ll 1$) $\langle\hat H_M\rangle$ is constant and the 
OS classical trajectories for dust are recovered. 
This in turn guarantees {\em a posteriori} the identification
of the quantized scalar field (mode) with dust.
\par
In the present approach neither a scalar field nor other models for 
the dust are used, leaving room for future developments and possible 
overlappings with the cited papers.
On the other side the matching conditions are implemented from the
onset and need not to be imposed by hand later.
Further, the ADM surface term is also included from the start
and correspondingly the ADM energy is not {\em a priori} fixed.
\par
A further improvement of our approach with respect to the cited 
papers is that our reduced actions $I_{r_s}^\epsilon$ describe 
larger classes of OS solutions and the extended action $I^e$ covers 
all the OS solutions. 
This is no real improvement for the classical theory, since there one 
only needs to describe one trajectory at a time (not even one class),
but is important in the quantized theory, as can be seen already from 
the following semi-classical argument.
Since dust proper energy is not {\em a priori} conserved,
one could have a non-vanishing probability for the transition from a 
state with $H_M$ corresponding to a certain (semi-)classical trajectory 
in $[E,m]$ to a different $\bar H_M$, with $m$ held constant.
From (\ref{prop}) and the first equation in (\ref{init}) one can then 
interpret this as a quantum transition to a (semi-classical) trajectory 
with different comoving radius $\bar r_s$. 
The latter trajectory belongs to a different class $[\bar E,m]$ and is 
described by the super-Hamiltonian $H^\epsilon_{\bar r_s}$ such that 
Eq.~(\ref{prop}) still holds for the same value $m$ of the ADM energy.
We note in passing that such change in $r_s$ represents a ``motion''
of the surface of the sphere with respect to the comoving frame and
would thus correspond to the presence of effective (quantum) pressure
(see {\em e.g.} \cite{narlikar}).
If, according to Ref.~ \cite{cfv}, $\langle\hat H_M\rangle$ 
steadily increases in time, then $E$ converges to $1$ and the class 
$[E=1]$ becomes a sort of {\em quantum attractor} in the space of 
semi-classical states of the OS model.
\par
In a full quantum context one can go even further and imagine a state 
which is a superposition of several $m$-eigenstates (belonging to the 
same $[E]$ or not).
Such a state is allowed by the indetermination principle applied to the 
canonical variables $m$ and $p$ and would represent a superposition
of space-times where asymptotic ($r\gg r_s$) trajectories of test 
particles move according to a ``superposition'' of Newton laws.
\section{Conclusions}
\label{conclude}
In these notes we have carried on a derivation of canonical
actions for the OS model which takes into account from the onset
the matching conditions at the surface of the sphere and the ADM 
surface term.
This allowed us to identify the range of applicability of the
actions already found in the literature and extend them to cover
the full space of OS solutions. 
\par
In particular we have shown that the full phase space of the model
is naturally parted into classes defined by the geodesic
energy of the surface of the sphere which uniquely specifies the
inner geometry. 
We have then introduced actions describing classes of solutions 
which are larger than the classes obtained by fixing the ADM energy
and the geodesic energy, as we have shown that was (implicitly) done 
in previous papers \cite{lund,peleg,cv}.
\par
We think the above results are useful for a better understanding 
of the quantization of the OS model as a constrained dynamical 
system.
\acknowledgments
I would like to thank P. Hajicek for suggesting the problem and many
useful discussions and the Institute for Theoretical Physics of the 
University of Bern for the kind hospitality during February 1998.
%

%

\begin{references}
\bibitem{co_orto}
In principle one can define a comoving frame which is not time-orthogonal,
but this would lead to unnecessary complications, such as the need for a 
non-vanishing radial lapse function.
\end{references}
\end{document}